*Institute of Metal Physics, Academy of Sciences of the Ukrainian SSR, Kiev*

# The Plasma Spectrum Effect on Acoustic Wave Amplification by Carrier Drift in Semiconductors

By

Yu. V. Kornyushin

The effect of plasma oscillations on acoustic wave amplification by carrier drift in semiconductors is discussed. A continuum theory is developed on the basis of hydrodynamical equations for the conduction electrons, the equations for the elastic vibrations of the crystal, and the Poisson equation. It is shown that if the plasma frequency $\omega_{p0}$ is much lower than the reciprocal value of the electron relaxation time for the scattering on phonons and impurities, $\tau^{-1}$, the results are in agreement with available calculations based on the equation for electrical conductivity and diffusion. However, in the case $\omega_{p0}\,\tau \gg 1$ the amplification factor is multiplied by $2\,\omega_p\tau$ compared to the value predicted by those calculations, while the optimum drift velocity in this case is $\omega_p\tau$ times smaller.

В работе рассматривается влияние спектра плазменных колебаний на усиление акустических волн дрейфом носителей в полупроводниках. Континуальная теория строится на основании гидродинамических уравнений для электронов проводимости, уравнений теории упругости движения кристалла и уравнения Пуассона. Показано, что если плазменная частота $\omega_{p0}$ гораздо меньше обратного времени релаксации электронов при рассеянии на фононах и примесях $\tau^{-1}$, то результаты совпадают с имеющимися в литературе, и полученными на основании уравнения электропроводности-диффузии. В противоположном случае, когда $\omega_{p0}\,\tau \gg 1$, показано, что при менее жестком требовании к дрейфовой скорости, чем в существовавших ранее теориях, коэффициент усиления увеличивается в $2\,\omega_p\tau$ раз по сравнению с величиной, предсказываемой в теориях, построенных на основании уравнения электропроводности-диффузии.

## 1. Introduction

In recent years the investigation of acoustic effects in semiconductors and particularly of acoustic wave amplification by carrier drift has become a separate branch of the physics of semiconductors.

The possibility of acoustic wave amplification in crystals was theoretically predicted in 1956 by Tolpygo and Uritskii [1] and by Weinreich [2]. Hypersound amplification was first obtained experimentally by Hutson et al. [3] in piezoelectric crystals. Pomerantz [4] was the first who observed sound amplification in n-Ge using electric fields. He compared the amplification factor obtained with the factor deduced from the theory of Weinreich et al. [5]. The amplification factor observed experimentally turned out to be about 50 times lower than the calculated one.

At present both the microscopic theory of acousto-electrical effects (see for example Gantsevich and Gurevich [6] and Gantsevich [7]) and the continuum theory of acoustic wave amplification are developed.

The continuum theory of amplification distinguishes the following mechanisms of electron–phonon interaction:

1. The deformation potential. This mechanism was studied by Weinreich [2], Spector [8], Kazarinov and Skobov [9], and others.

2. The piezoelectrical interaction (see for example Pekar [10]).

3. The electron–phonon interaction proportional to the external applied field and defined by the dependence on deformations of the dielectric constant. Such an interaction was proposed by Pekar [10]. It is essential in crystals with a high value of the dielectric constant.

4. Lattice dragging by the drift of conduction electrons, which strength was studied in the work of Pekar and Tsekvava [11].

5. It is also possible to consider the electron–phonon interaction due to the dependence of the mobility of the current carriers on the deformation, or, equivalently, to take into account the dependence of the relaxation time $\tau$ and the effective mass $m$ on the deformation.

The essence of continuum theories of acoustic wave amplification consists in the simultaneous solution of uniform equations: the continuity equation for the current, the Poisson equation for the electrostatic potential, and the elasticity theory equations for medium displacements. If the equations mentioned do not separate into a sum of independent equations for unknown functions (the charge density, the electric potential, and the medium displacements) the condition for their solubility will be a secular equation which gives the wave dispersion law containing imaginary terms responsible for amplification.

The current in the continuum theory of acoustic wave amplification is given by the electrical conductivity and diffusion equation. This excludes automatically the possibility from the theory to take into account plasma oscillations of current carriers. In the majority of microscopic theories of acoustic wave amplification plasma oscillations are not considered.

It appears that the spectrum of plasma oscillations may very significantly affect the amplification factor (particularly under experimental conditions). The method developed in the present work allows to take into account the plasma oscillation spectrum within the scope of the continuum theory of acoustic wave amplification. The method uses a hydrodynamical description of the current carrier system. As an electron–phonon interaction the deformation potential is taken.

## 2. Drift and Plasma Waves

Drift waves were considered by many authors, e.g. Ridley [12] and Gurevich and Ioffe [13]. Drift waves in a one-zone model follow from a synchronous solution of the continuity equation for the current density,

$$e\,\dot{n} + (\triangledown, \boldsymbol{I}) = 0 , \tag{1}$$

where $e$ is the charge of current carriers and $n$ the carrier concentration, and the Poisson equation

$$(\triangledown, \boldsymbol{E}) = \frac{4\pi}{\varepsilon}\, e\,(n - n_0) , \tag{2}$$

where $\varepsilon$ is the dielectric constant, $\boldsymbol{E}$ the electric field, and $n_0$ the average concentration of the current carriers. It is assumed that the current density is given by the electrical conductivity and diffusion equation

$$\boldsymbol{I} = e\,\mu_0\,n\,\boldsymbol{E} - e\,D\,\nabla\,n\,, \tag{3}$$

where $\mu_0$ is the mobility of the current carriers in the zone and $D$ is their diffusion constant. From (1) to (3) one can obtain the drift wave dispersion law

$$\omega = V_{0\|}\,k - i\,\tau_k^{-1}\,, \qquad V_{0\|} = \mu_0\,\frac{(\boldsymbol{E}_0,\,\boldsymbol{k})}{k}\,, \tag{4}$$

where $\tau_k^{-1} = (4\,\pi\,e\,\mu_0\,n_0)/\varepsilon + D\,k^2$, $\boldsymbol{E}_0$ is the external homogeneous field, $\boldsymbol{k}$ the wave vector, and $\omega$ the frequency. Clearly drift waves are present if $V_{0\|}\,k\,\tau_k \gg 1$.

Plasma waves in a hydrodynamical approximation when external homogeneous fields are present were investigated by Bass et al. [14] and Kikvidze et al. [15].

Let us write the Poisson equation in the form

$$\triangle\,\varphi + m\,\omega_{p0}^2\,\frac{n - n_0}{e\,n_0} = 0\,, \qquad \omega_{p0}^2 = \frac{4\,\pi\,e^2\,n_0}{\varepsilon\,m}\,, \tag{5}$$

where $\omega_{p0}$ is the plasma frequency. The linearized continuity equation has the form

$$\dot{n} + n_0\,(\nabla,\,\boldsymbol{v}) + (\boldsymbol{V}_0,\,\nabla\,n) = 0\,. \tag{6}$$

The linearized Euler equation for the current carrier velocity may be written

$$\dot{\boldsymbol{v}} + (\boldsymbol{V}_0,\,\nabla)\,\boldsymbol{v} + \frac{\boldsymbol{v}}{\tau} + \frac{e}{m}\,\nabla\,\varphi + \alpha\,\frac{\omega_{p0}^2}{g^2\,n_0}\,\nabla\,n = 0\,, \tag{7}$$

where $\tau$ is the carrier relaxation time for collisions with the lattice, $\mu_0 = e\,\tau/m$ the mobility, $g^{-1}$ the screening radius of Debye-Hückel or Thomas-Fermi depending on the degree of degeneracy of the current carriers. $\alpha\,(\omega_{p0}^2/g^2\,n_0)\,\nabla\,n$ is the term describing the electron-gas pressure. When the oscillation frequency of the system is much higher than the collision frequency $\tau^{-1}$ then, as was shown in [16], the oscillations of the electron system are essentially one-dimensional and adiabatic ($\alpha = 3$ in the case of Boltzman statistics, and $\alpha = 1.8$ for a degenerated electron gas). In the opposite case the electron motion is not one-dimensional but isothermal. This corresponds to $\alpha = 1$. From (5) to (7) one gets the dispersion law of plasma waves,

$$\omega_{1,2} = V_{0\|}\,k - \frac{i}{2\,\tau} \pm \sqrt{\omega_p^2 - \frac{1}{4\,\tau^2}}\,, \qquad \omega_p^2 = \omega_{p0}^2\left(1 + \alpha\,\frac{k^2}{g^2}\right). \tag{8}$$

The result (8) for $V_0 = 0$ and $\tau^{-1} = 0$ coincides with the corresponding result obtained from quantum theory [17].

We shall further consider the case of low-frequency oscillations $\omega\,\tau \ll 1$ which corresponds to $\alpha = 1$.

With arbitrary parameter values the dispersion law of plasma waves (8) is not quite similar to the dispersion law of drift waves (4). However at $4\,\omega_{p0}^2\,\tau^2 \ll 1$, i.e. at those values of the relaxation time when it is incorrect to speak about plasma waves, we get from (8)

$$\omega_1 = V_{0\parallel}\, k - i\, \tau_k^{-1}\,, \qquad \omega_2 = V_{0\parallel}\, k - i\left(\frac{1}{\tau} - \frac{1}{\tau_k}\right). \qquad (9)$$

The second wave attenuates much stronger than the first so that it may neglected.

From the comparison of drift and plasma waves it can be seen that the conductivity and diffusion equations may be applied only in the case when $4\,\omega_{p0}^2\, \tau^2 \ll 1$, i.e. when plasma oscillations are absent. In other cases more general hydrodynamical equations should be used and this markedly influences in particular the expression for the amplification factor of acoustic waves.

## 3. Hydrodynamical Theory of Acoustic Wave Amplification

To investigate acoustic wave amplification it is necessary to add to the Poisson equation (5) and to the continuity equation (6) the Euler equation for the electron velocity,

$$\dot{\boldsymbol{v}} + (\boldsymbol{V_0},\, \nabla)\, \boldsymbol{v} + \frac{\boldsymbol{v}}{\tau} + \frac{e}{m}\, \nabla\, \varphi + \frac{\omega_{p0}^2}{g^2\, n_0}\, \nabla\, n + \frac{e}{m}\, C\, \nabla\, (\nabla,\, \boldsymbol{u}) = 0\,, \quad (10)$$

and the equation for medium displacement from the elasticity theory,

$$\gamma\, \ddot{\boldsymbol{u}} - (\lambda + \mu)\, \nabla\, (\nabla,\, \boldsymbol{u}) - \mu\, \triangle \boldsymbol{u} - e\, C\, \nabla\, n = 0\,. \qquad (11)$$

In (10) and (11) $\boldsymbol{u}$ is the vector of medium displacement, $C$ is the deformation potential constant, $\gamma$ the crystal density, $\lambda$ and $\mu$ are the Lamé coefficients.

For the case $\boldsymbol{E_0} \parallel \boldsymbol{k}$ (5), (6), (10), and (11) lead to the following dispersion equation for the longitudinal waves:

$$\left[(\omega - V_0\, k)^2 + i\, \frac{\omega - V_0\, k}{\tau} - \omega_p^2\right][(\lambda + 2\,\mu)\, k^2 - \gamma\, \omega^2] + \frac{e^2\, C^2\, k^4\, n_0}{m} = 0\,. \quad (12)$$

From (12) for a quasi-acoustic wave at a given real frequency we get

$$\frac{\mathrm{Im}\, \Delta k}{k} = \frac{\varepsilon}{8\,\pi}\, \frac{C^2\, k^2}{\lambda + 2\,\mu}\, \frac{g^2}{k^2 + g^2}\, \frac{(\omega - V_0\, k)\, \tau_k}{(\omega - V_0\, k)^2\, \tau_k^2 + [1 - (\omega - V_0\, k)^2\, \tau_k^2\, \omega_p^2\, \tau^2]^2}\,. \quad (13)$$

In the case $\omega_p\, \tau \ll 1$, as already mentioned in Section 2, the plasma waves are of drift type, and according to this from (13) we obtain the usual expression for the amplification factor which is in agreement with the corresponding results published in [2, 10]. The optimum value of the drift velocity in this case is defined by

$$V_0 = S_\parallel + (k\, \tau_k)^{-1}\,, \qquad (14)$$

where $S_\parallel$ is the velocity of the longitudinal acoustic waves under consideration. The last fraction in the right-hand side of (13) attains a value equal to $-1/2$ when (14) is satisfied.

If, however, for example $\varepsilon = 3$, $m = 10^{-27}$ g, $\mu_0 = 500$ cm$^2$ V$^{-1}$ s$^{-1}$, and $n_0 = 10^{17}$ cm$^{-3}$, then we get $\omega_{p0}\, \tau = 3$ and under the condition (14) the last fraction in the right-hand side of (13) has the value $-1/65$. This means that in this case the amplification factor will be 32 times smaller than the value obtained from formulas existing in literature. This may serve as a possible explanation of disagreement between theoretical and experimental data mentioned in Section 1.

Now let us consider the case $\omega_p \tau \gg 1$. It is usually discussed in works on coherent amplification of spin waves by plasma beams [18, 19]. In this case, at $\omega + \omega_p = V_0 k$, the amplification factor has the maximum value defined by the formula

$$\frac{\mathrm{Im}\,\Delta k}{k} = -\frac{\varepsilon}{8\,\pi} \frac{C^2\,k^2}{\lambda + 2\,\mu} \frac{g^2}{k^2 + g^2}\,\omega_p\,\tau\,,\tag{15}$$

which is $2\,\omega_p\,\tau$ times larger than in the preceding case. The condition for optimum amplification may be written in another form

$$V_0 = S_{\parallel} + \frac{\omega_p}{k} = S_{\parallel} + S_p\,.\tag{16}$$

From this equation one may clearly recognize the difference between the condition (16) and those published in literature for the optimum amplification (14). In our case, when $\omega_p \tau \gg 1$ the phase velocity of plasma waves $S_p$ is $\omega_p \tau$ times lower than $(k\,\tau_k)^{-1}$, i.e. in this case condition (16) requires a much lower drift velocity than condition (14).

If the condition $\omega_p \tau \gg 1$ is satisfied and $V_0 = S_{\parallel} - S_p$, then, according to (13), maximum absorption of ultrasound should be observed. At sufficiently high values of $\omega_p \tau$ and if condition (16) is fulfilled, the secular equation (12) has roots close together [20] (in the zero electron–phonon interaction approximation). In this case (13) and (15) are incorrect. We shall not consider this case in the present work because it is difficult to realize this case experimentally. In fact, in practice as a rule we have $\omega \lesssim 10^{10}$ s$^{-1}$ and $\tau \lesssim 5 \times 10^{-11}$ s. With these values of the parameters the approximation of separate roots in (12) works well.

The situation of close roots ($\tau^{-1} = 0$) was considered in the work of Solymar [21].

## 4. Conclusions

The hydrodynamical theory of acoustic wave amplification by carrier drift was considered. In it the electron–phonon interaction is taken into account only through the deformation potential. If other mechanisms of electron–phonon interaction should be taken into account, $C^2\,k^2$ in (13) will perhaps be replaced by other expressions describing the electron–phonon interaction. However, it was the aim of this work to study the effects associated with the presence of the plasma oscillation spectrum. These effects are contained in (13) in the factor

$$f = \frac{(\omega - V_0\,k)\,\tau_k}{(\omega - V_0\,k)^2\,\tau_k^2 + [1 - (\omega - V_0\,k)^2\,\tau_k^2\,\omega_p^2\,\tau^2]^2} =$$
$$= \frac{(\omega - V_0\,k)\,\tau_k}{(\omega - V_0\,k)^2\,\tau_k^2 + \left[1 - \left(\dfrac{\omega - V_0\,k}{\omega_p}\right)^2\right]^2}\,.\tag{17}$$

Equation (17) shows that for any frequency of the acoustic waves the parameter $\omega_p \tau$ may serve as a criterion which decides whether it may be necessary to take into account plasma effects. It is noteworthy that these effects are also significant far from the plasma acoustic resonance. It was already mentioned in Section 3 that at $\omega_p \tau \ll 1$ plasma effects are negligible and the results coincide with those in the literature. At $\omega_p \tau \gg 1$, the amplification factor is to be multiplied by $2\omega_p \tau$ if we fulfil the optimum drift condition (16). This condition requires a lower drift velocity than condition (14) derived from the drift theory of acoustic wave amplification.

However, it should be mentioned that the higher the value $\omega_p \tau$ the more the factor (17) is sensitive to deviations of the drift velocity from the optimum value (16). Thus for example at $\omega_p \tau = 100$ a deviation of the drift velocity by 10% from the phase velocity of the plasma waves results in a decrease of the factor (17) by four times. At $\omega_p \tau \to \infty$ the slightest deviation makes the factor (17) to drop to zero. Hence the optimum condition (16) may be called a resonance condition. When the value $\omega_p \tau$ is arbitrary the condition for optimum drift velocity becomes more complicated and is omitted here. One can see that at optimum drift velocity the amplification factor increases with increasing $\omega_p \tau$.

The theory developed does not take into account heating of current carriers in the electric field and hence is applicable only for electric fields not too high. The microscopic theory [6, 7] points out criterions for the heating of current carriers.